\documentclass[aps,prb,twocolumn,superscriptaddress,10pt]{revtex4-1}
\usepackage[utf8]{inputenc}
\usepackage{libertine}
\usepackage[libertine]{newtxmath}
\usepackage{mathtools,graphicx,microtype,bm}
\usepackage[cal=boondoxo,frak=boondox]{mathalfa}
\usepackage[colorlinks=true,allcolors=blue]{hyperref}

\begin{document}

\title{Long-range perturbation of helical edge states by nonmagnetic defects in two-dimensional topological insulators}

\author{Vladimir A.\ Sablikov}
\email[E-mail:]{sablikov@gmail.com}
\affiliation{Kotel'nikov Institute of Radio Engineering and Electronics,
Russian Academy of Sciences, Fryazino, Moscow District, 141190, Russia}

\begin{abstract}
We study the electronic states that are formed due to the tunnel coupling between helical edge states (HESs) and bound states of nonmagnetic point defects in two-dimensional topological insulators in the general case of broken axial spin symmetry. It is found that the coupling of HESs and a single defect leads to the formation of composite HESs composed of the bound states and a set of the conventional HESs. Their spectral density near the defect has a resonance shifted relative to the energy level of the bound state. But of most importance is a long-range perturbation of the HESs around the defect, which is a cloud consisting of both Kramers partners of conventional edge states. Therefore each of the composite HESs contains both the right- and left-moving conventional HESs. The amplitude of this perturbation decreases inversely with the distance from the defect. In a system of many defects, this perturbation leads to a long-range coupling between bound states of different defects mediated by the HESs and causes amazing effects. We study these effects for a two-defect system where the proposed mechanism of indirect coupling leads to a splitting of the resonances of isolated defects even if the distance between them is very large. As a result an asymmetric structure of two-peak resonance arises that very unusually changes with the distance between the defects.
\end{abstract}

\maketitle

\section{Introduction}

Helical edge states (HESs) with a gapless spectrum are a hallmark of quantum spin-Hall systems that have attracted much interest over a decade~\cite{RevModPhys.82.3045,RevModPhys.83.1057,Ren_2016} starting from pioneering prediction of HESs~\cite{PhysRevLett.95.226801,PhysRevLett.95.146802,PhysRevLett.96.106802} and their experimental observation~\cite{Konig766}. HESs are a pair of counterpropagating, spin-polarized channels located at the edges of the sample in which the spin of an electron is locked to its momentum. Since the system is symmetric with respect to the time reversal, HESs are a Kramers doublet and therefore elastic scattering of electrons in these states is impossible~\cite{PhysRevB.73.045322,RevModPhys.82.3045,RevModPhys.83.1057,Ren_2016,bernevig2013topological}. However, experiments show that in reality backscattering does occur in the absence of magnetic impurities and the rate of this process is much higher than might be expected~\cite{PhysRevLett.123.047701,PhysRevX.3.021003,GUSEV2019113701}. A physical mechanism that would explain quantitatively or at least qualitatively the observed suppression of conductivity has not yet been established, though it is evident that two factors are important: the presence of impurities or other structure defects and breaking the axial spin symmetry due to spin-orbit interaction (SOI). Difficulties in solving this problem are probably associated with a lack of understanding of how electrons in edge states interact with nonmagnetic defects. This motivated us to study in more detail the electronic structure and spin texture of the edge states coupled to isolated nonmagnetic defects in two-dimensional (2D) topological insulators (TIs) with broken axial spin symmetry. 

In the absence of axial spin symmetry, the spin is not a good quantum number, and the eigenstates are classified by their Kramers index that determines also the direction of motion. In the framework of a minimal four-band model, such as the model of Bernevig, Hughes, and Zhang (BHZ)~\cite{Bernevig1757}, they are described by fourth-order spinors with a certain set of two spin and two orbital (pseudospin) components. Such helical states are often called generic ones~\cite{PhysRevLett.108.156402}. 

In TIs with isolated defects, there are two types of generic states with the energy within the band gap. First, there are HESs that are localized near the edge and have a well defined projection of the momentum along it. They were widely studied in the literature~\cite{PhysRevLett.108.156402,PhysRevB.91.245112,PhysRevB.93.205431}. The second type of generic states is bound states induced by point defects and impurities. The presence of the bound states is important, since they are formed in almost any potential of the defect~\cite{Lu_2011,doi:10.1002/pssr.201409284,PhysRevB.92.085126}, unless it is too smooth or too small. For the systems with broken axial spin symmetry these types of states are not so widely studied. Their spectrum and spinor wave function were calculated only in some specific cases~\cite{SABLIKOV20161,PhysRevB.91.245112,doi:10.1063/1.5049717}. 

An interesting situation arises when a defect is located close to the edge, and the bound states interact with a continuum of edge states. In essence, this situation is similar to the configuration interaction of localized states with a continuum in the Fano–Anderson theory~\cite{PhysRev.124.1866,mahan2013many}. Previously, we showed that in this case new edge states are formed that flow around the defect and have a resonance of local density of states~\cite{PhysRevB.91.075412}, but we did not study their electronic structure and, most importantly, did not study these states under conditions of broken axial spin symmetry, when a strong change in their spin structure can be expected.

It is important that in systems with broken axial spin symmetry, the spinor structures of the edge and bound states are very different, if only because their spatial configurations are very different: Edge states move along a straight line while bound states are circular. Therefore we can expect that the coupling of these states will lead to a strong perturbation of the continuum of edge states. 

In this paper, we study the effects of the tunnel coupling between HESs and one or more defects, using the general restrictions imposed on the four-rank spinors of the edge and bound states by the time reversal symmetry. Specific calculations, where necessary, are performed within the framework of the BHZ model. We show that the coupling between conventional HESs and bound states leads to the formation of a Kramers doublet of propagating states, which are composed of the bound states and a wide set of the conventional HESs. The amplitude of these composite states in the vicinity of the defect has a certain resonant structure. The set of the HESs forms a cloud that extends far from the defect. In the case of several defects, a new mechanism of an {\em indirect} coupling between defects through the edge states appears, which can couple the defects at large distance, giving rise to significant changes in the structure of the resonances.

The structure of the paper is as follows. In Sec.~\ref{Sec2} we introduce the generic HESs and bound states and present a theory of the composite HESs in the case where there is a single defect. Section~\ref{Sec3} is devoted to the composite HESs in a system of two defects. Here we find the wave functions of the composite HESs, introduce the notion of the indirect coupling between defects mediated by edge states, study the spectrum of the wave-function amplitude and discuss the effects of the indirect coupling. Section~\ref{Sec4} summarizes main results. In Appendix~\ref{App1} we derive an expression for the tunneling Hamiltonian coupling the HESs and bound states. Appendix~\ref{App2} contains details of the calculation of the wave function for the system of two defects coupled to HESs. 

\section{Helical edge states coupled to a defect}\label{Sec2}

We begin with a short reminder of how HESs and bound states are described in the 2D TIs with broken axial spin symmetry within a four-band model, such as the BHZ model. 

\subsection{Helical edge states}\label{HESs}
HESs were studied in recent years~\cite{PhysRevLett.108.156402,PhysRevB.91.245112,PhysRevB.93.205431,Durnev_2018} and their main properties were understood. There are two sets of counterpropagating HESs 
\begin{equation}
\Psi_{k,\sigma}(x,y)=\frac{1}{\sqrt{L}}\widetilde{\Psi}_{k,\sigma}(y)\exp{(i k x-i\varepsilon_{k,\sigma}t)},\end{equation}
labeled by the momentum $k$ and the Kramers index $\sigma=\pm$, which also indicates the propagation direction. Here $x$ is the coordinate along the edge, $y>0$ is normal coordinate, $\varepsilon_{k,\sigma}$ is the energy, $\widetilde{\Psi}_{k,\sigma}$ is a four-rank spinor describing the $y$-dependence of the wave function, and $L$ is a normalization length. 

The energies of the right- and left-moving HESs are related to each other due to the time reversal symmetry
\begin{equation}
\varepsilon_{k,\sigma}=\varepsilon_{-k,-\sigma}.
\end{equation}
The dependence of the energy on $k$ is very close to linear, $\varepsilon_{k,\sigma}\approx \sigma v k$, with $v$ being velocity. The spinors $\widetilde{\Psi}_{k,\sigma}(y)$ with opposite $\sigma$ are also related as components of the Kramers doublet. They can be written in the form: 
\begin{equation}
\widetilde{\Psi}_{k,+}(y)=\widetilde{\Psi}_{k}(y) \equiv
\begin{pmatrix}
\psi_{1,k}(y) \\ \psi_{2,k}(y)  \\ \psi_{3,k}(y) \\ \psi_{4,k}(y)
\end{pmatrix},\quad
\widetilde{\Psi}_{k,-}(y)=
\begin{pmatrix}
-\psi_{3,-k}^*(y) \\ -\psi_{4,-k}^*(y)\,  \\ \psi_{1,-k}^*(y) \\ \psi_{2,-k}^*(y)
\end{pmatrix}.
\end{equation}

In the literature, a simplified model is often used in which the four-component wave function is effectively replaced by a two-component one describing a state with spin rotated by an angle that depends on the momentum~\cite{PhysRevLett.108.156402,PhysRevB.91.245112,PhysRevB.93.205431}. In this way, the essential features of the HESs are well captured for weak SOI, if the HESs are considered as one dimensional. In our case, this approach is not constructive, since it ignores the dependence of the wave function on the normal coordinate $y$, while it is important for us to calculate the overlap integrals of the edge and bound states. In addition, the bound states, in any case, are four-rank spinors. 

The $y$ dependence of the wave function is presented by a sum of four exponentially decaying terms, in accordance with the order of the differential equations describing $\widetilde{\Psi}_{k,\sigma}(y)$: 
\begin{equation}
\psi_{i,k}(y)=\sum \limits_{j=1,4} C_{i,j}(\varepsilon,k) e^{-\kappa_j y}\,,
\label{edge-spinor-comp}
\end{equation}
where $\kappa_j(\varepsilon,k)$ is a complex value with positive real part. The coefficients $C_{i,j}$ are determined by a system of four linear equations, which follows from the corresponding Schr\"odinger equation. The determinant of this system gives the dispersion equation defining $\varepsilon_{k,\sigma}$ for the HESs. 

We use this procedure for specific numerical calculations of the wave functions and matrix elements.  

\subsection{Bound states}
Bound states in the presence of SOI are studied much less than HESs. They were studied mainly by numerical calculations for a number of specific situations, such as defects with a short-range potential~\cite{SABLIKOV20161}, a Coulomb impurity in a quantum dot~\cite{doi:10.1063/1.5049717}, and a quantum disk of large radius~\cite{PhysRevB.91.245112}. Nevertheless, it is clear that there is a set of states $\Phi_{n,\lambda}$ characterized by a quantum number $n$, which indicates the energy level, and the Kramers index $\lambda=\pm$. Further in this paper, for simplicity, we restrict ourselves to one (ground) energy level $\varepsilon_0$ and the corresponding Kramers doublet of states which can be written in the form
\begin{equation}
\Phi_+(r,\varphi)=
\begin{pmatrix}
\phi_{1}(r) \\ i\phi_{2}(r)e^{-i\varphi} \\ i\phi_{3}(r)e^{-i\varphi} \\ \phi_{4}(r)
\end{pmatrix},\quad
\Phi_-(r,\varphi)=
\begin{pmatrix}
i\phi_{3}^*(r)e^{i\varphi} \\ -\phi_{4}^*(r) \\ \phi_{1}^*(r) \\ -i\phi_{2}^*(r)e^{i\varphi}
\end{pmatrix},
\end{equation}
where $r$ and $\varphi$ are polar coordinates with the center at the defect. These wave functions obviously describe states circulating clockwise and counterclockwise around the defect. The functions $\phi_i(r)$ are defined by straightforward solution of the Schr\"odinger equation, which can be performed numerically. Such calculations will be required in what follows for quantitative estimates. They will be carried out for a defect with a short-range potential using the previously developed method~\cite{doi:10.1002/pssr.201409284,PhysRevB.91.075412,SABLIKOV20161}.

\subsection{Helical edge states coupled to a defect}\label{HESDs}

\begin{figure}
\includegraphics[width=0.9\linewidth]{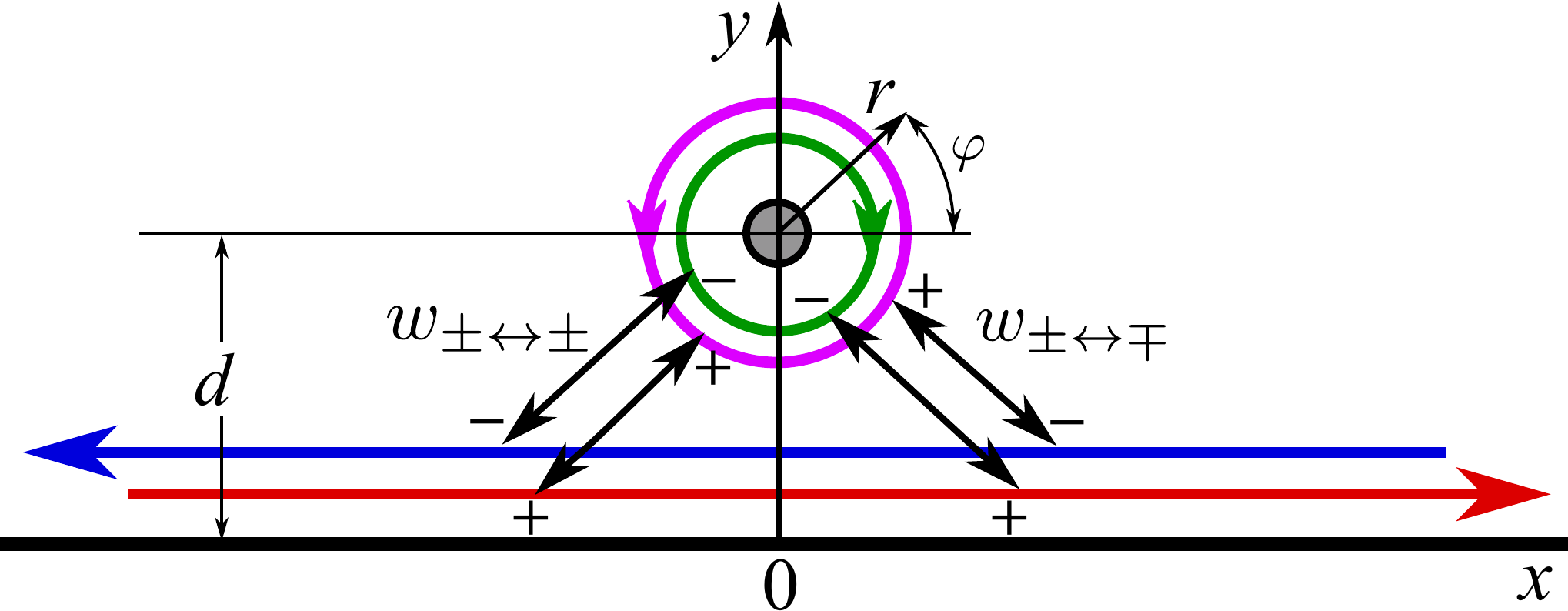}
 \caption{A point defect coupled to HESs and the electron transitions mixing the HESs and the bound states.} 
\label{fig_1}
\end{figure}

When the HESs are coupled to a defect located at some distance from the edge, the total system can be described by the tunneling Hamiltonian
\begin{equation}
H\!=\!\sum\limits_{k,\sigma}|k,\sigma\rangle \varepsilon_{k,\sigma}\langle k,\sigma| + \sum\limits_{\lambda}|\lambda\rangle \varepsilon_0\langle \lambda| + \sum\limits_{k,\sigma,\lambda}\!(|k,\sigma\rangle w_{k,\sigma;\lambda}\langle \lambda|+ h.c.)\,,
\label{HybridHamiltonian}
\end{equation}
where the first term is the HES Hamiltonian, the second term is the Hamiltonian of the bound states with the energy $\varepsilon_0$, and the third term is the Bardeen’s tunneling Hamiltonian. The sketch of a defect coupled to HESs and tunnel transitions are shown in Fig.~\ref{fig_1}. 

\subsubsection{Tunneling matrix}

In contrast to the case when the $z$-projection of the spin is well defined, the tunnel matrix is not diagonal in the Kramers indices $\sigma$ and $\lambda$. Therefore, the tunneling Hamiltonian mixes the right and left-moving HESs with both Kramers partners of the bound states. The tunneling matrix $w_{k,\sigma;\lambda}$ satisfies the relation
\begin{equation}
w_{k,\sigma;\lambda}=\sigma\lambda w_{-k,-\sigma;-\lambda}^*\,,
\label{w_symmetry}
\end{equation}
which follows from the time reversal symmetry.

In what follows, in addition to the general relation~(\ref{w_symmetry}), we will need a more detailed idea of how the matrix $w_{k,\sigma;\lambda}$ depends on $k$. This information can be obtained using the explicit form of the tunneling Hamiltonian. We have shown that, under fairly general assumptions, the tunneling Hamiltonian coupling the HESs and the defect coincides, up to sign, with the bulk Hamiltonian of the 2D TI. The proof of this statement is given in Appendix~\ref{App1}. 

As a model for specific calculations we will use the BHZ model~\cite{Bernevig1757} with SOI caused by the bulk inversion asymmetry~\cite{doi:10.1143/JPSJ.77.031007}. Therefore the matrix elements $w_{k,\sigma;\lambda}$ will be calculated using the BHZ Hamiltonian. Details of the model and calculation method are also given in Appendix~\ref {App1}.

It is clear from Eq.~(\ref{w_symmetry}) that only two components of the tunneling matrix, $w_{k,+;+}$ and $w_{k,+;-}$, are independent. They describe the tunnel transitions between one of the Kramers components of the HESs and the components of the Kramers doublet of the bound states. In the limiting case of the weak SOI ($\Delta \ll |M|$, with $M$ being the mass term in the BHZ model and $\Delta$ being the SOI parameter), one can roughly say that the matrix element $w_{k,+;+}$ describes the transitions with the same spin, and $w_{k,+;-}$ describes the spin flip transitions. We have studied the matrix elements $w_{k,+;+}$ and $w_{k,+;-}$ as a function of $k$. 

The main parameters of the model, which largely determine the spin-flip transitions, are the SOI parameter $\Delta /|M|$ and the parameter $A$ of the band hybridization, which is also normalized $a=A/\sqrt{|BM|}$. In many cases, the parameter $a$ plays the essential role since it determines the edge state velocity and one of the two lengths of the edge-state penetration deep into the sample.

The calculations show that the $k$ dependence of the tunneling matrix elements significantly varies with position of the defect relative to the edge, $d$. This is because the different components of the spinors $\widetilde{\Psi}_{k,\sigma}$ and $\Phi_{\lambda}$ vary with the coordinate $y$ differently. Nevertheless, there is a general pattern: Matrix elements increase when the energy of the HESs $\varepsilon$ approaches the edges of the gap due to an increase in the length of the HES penetration into the bulk. But when the energy enters the band, the matrix elements fall sharply, because the penetration length diverges and the HES disappears. This general regularity can be significantly distorted by an asymmetry of $w_{k,+;\pm}$ with respect to the sign of $k$, which appears because of the $k$ dependence of the spinor components. The asymmetry is the strongest for spin-flip transitions. 

\begin{figure}
\includegraphics[width=0.9\linewidth]{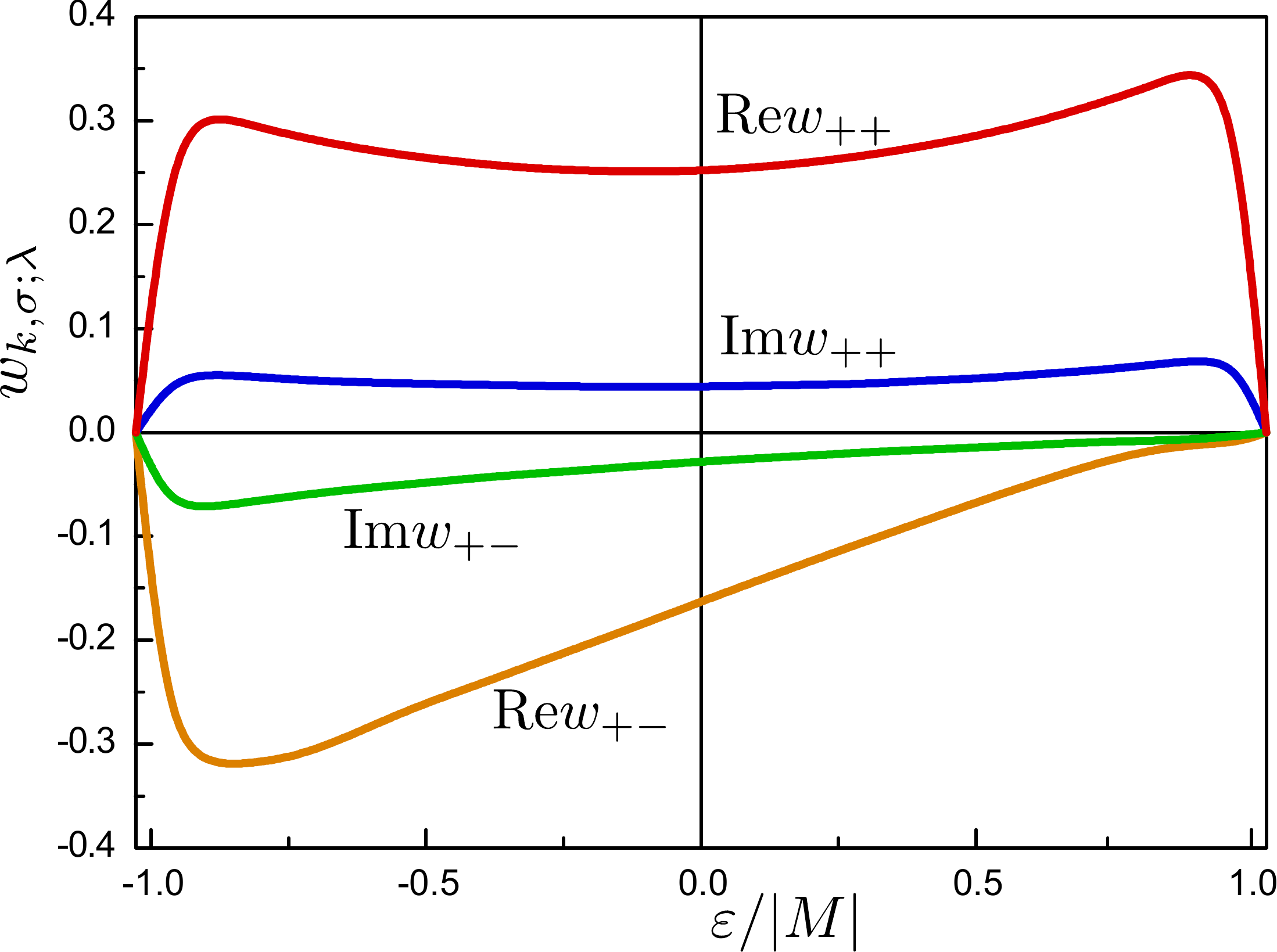}
 \caption{Tunneling matrix elements $w_{k,+;+}$ and $w_{k,+;-}$ as functions of the HES energy. The matrix elements are normalized to $|M|$ and shown without the normalization length factor $L^{-1/2}$. Numerical parameters used in the calculations are $\Delta=0.3 |M|$, $a=5$, $d = 12\sqrt{B/M}$.} 
 \label{fig_2}
\end{figure}

The results of the calculations are illustrated in Fig.~\ref{fig_2} for the model parameter $a=5$, which is close to that of HgTe/CdHgTe heterostructures, and $\Delta=0.3|M|$, which is a reasonable theoretical estimate of $\Delta$~\cite{WINKLER20122096,PhysRevB.93.075434}. The distance from the defect to the edge is $d=12\sqrt{B/M}$, where $\sqrt{|B/M|}$ is the characteristic length of the BHZ model, which characterizes also the localization of the bound states. The energy level of bound states depends on the potential of the defect and can be located anywhere inside the band gap. The results shown in Fig.~\ref{fig_2} are obtained for the energy level near the Dirac point.

It is seen that both components of the tunneling matrix are of the same order of magnitude and hence the Kramers doublets of the edge and bound states are very effectively mixed. Of course, such strong mixing occurs when the SOI is strong. In the first approximation, the mixing matrix element is linear in $\Delta$. 

\subsubsection{Wave functions}
Eigenfunction $\Psi$ of the Hamiltonian~(\ref{HybridHamiltonian}) can be constructed in the basis of the HESs and the bound states
\begin{equation}
\Psi=\sum\limits_{k',\sigma'}A_{k',\sigma'}|k',\sigma'\rangle + \sum\limits_{\lambda'}B_{\lambda'}|\lambda'\rangle\,.
\end{equation}
From the stationary Schr\"odinger equation $H\Psi=E\Psi$ we get the following equation system for the coefficients $A_{k,\sigma}$ and $B_{\lambda}$:
\begin{align}
\varepsilon_{k,\sigma}A_{k,\sigma}+\sum\limits_{\lambda'}w_{k,\sigma;\lambda'}B_{\lambda'} &=E\, A_{k,\sigma},\\
\varepsilon_0 B_{\lambda} +\sum\limits_{k',\sigma'}w_{k',\sigma';\lambda}^*A_{k',\sigma'} &=E\, B_{\lambda}.
\end{align}

The equations are solved by the methods of the theory of the Fano-Anderson model~\cite{PhysRev.124.1866,mahan2013many}. In this way we arrive at the following two wave functions:
\begin{equation}
\Psi_{\lambda}=\Phi_{\lambda}+\frac{1}{v}\sum\limits_{\sigma'}\sum\limits_{k'}\mathcal{P}\frac{w_{k',\sigma';\lambda}}{\mathcal{K}-\sigma'k'}\Psi_{k',\sigma'}+Z_{\mathcal{K}}\sum\limits_{\sigma'}w_{\sigma' \mathcal{K},\sigma';\lambda}\, \Psi_{\sigma'\mathcal{K},\sigma'},
\label{Psi_lambda}
\end{equation}
where $\mathcal{P}$ denotes the principal value, $\mathcal{K}=E/v$ is the wave vector of the edge states with the energy $E$ (for simplicity we put $\hbar=1$ hereinafter), and
\begin{equation}
Z_{\mathcal{K}}=\frac{E-\varepsilon_0-\Sigma_{\mathcal{K}}}{F_{\mathcal{K}}}.
\label{Z_K_1}
\end{equation}
Here the self-energy function $\Sigma_{\mathcal{K}}$ is 
\begin{equation}
\Sigma_{\mathcal{K}}=\frac{1}{v}\sum\limits_{k'}\mathcal{P}\frac{|w_{k',+;+}|^2+|w_{k',+;-}|^2}{\mathcal{K}-k'}\,,
\label{selfenergy0}
\end{equation}
and
\begin{equation}
F_{\mathcal{K}}=|w_{\mathcal{K},+;+}|^2+|w_{\mathcal{K},+;-}|^2\,.
\end{equation}

We expect that there should be two wave functions of the edge states coupled to the defect $\Psi_{\mathcal{K},R}$ and $\Psi_{\mathcal{K},L}$, corresponding right- and left-moving states. They satisfy the following conditions at infinity:
\begin{align}
&\Psi_{\mathcal{K},R}\Big|_{x\to \infty}  = Const\; \Psi_{\mathcal{K},+},
\label{b_conditionsR}\\
&\Psi_{\mathcal{K},L}\Big|_{x\to -\infty} = Const\; \Psi_{\mathcal{K},-}.
\label{b_conditionsL}
\end{align}
To satisfy these boundary conditions, the wave functions $\Psi_{\mathcal{K},R}$ and $\Psi_{\mathcal{K},L}$ are represented as a linear combination of the functions $\Psi_{\lambda}$:
\begin{equation}
\Psi_{\mathcal{K},R(L)}=\sum\limits_{\lambda'}B_{\lambda',R(L)}\Psi_{\lambda'}\,.
\end{equation}
The coefficients $B_{\lambda',R(L)}$ are easy to find from Eqs.~(\ref{b_conditionsR}) and (\ref{b_conditionsL}), and we get the following expressions for the right- and left-moving wave functions
\begin{multline}
\Psi_{\mathcal{K},R}=B_{\mathcal{K}}\Biggl\{w_{\mathcal{K},+;+}^*\Phi_++w_{\mathcal{K},+;-}^*\Phi_- +F_{\mathcal{K}}Z_{\mathcal{K}}\Psi_{\mathcal{K},+} \\+\dfrac{F_{\mathcal{K}}}{v}\sum\limits_{k'}\mathcal{P}\Biggl[\frac{\rho_1(\mathcal{K},k')}{\mathcal{K}-k'}\Psi_{k',+} +\frac{\rho_2(\mathcal{K},k')}{\mathcal{K}+k'}\Psi_{k',-}\Biggr] \Biggr\},
\label{RHESD}
\end{multline}
\begin{multline}
\Psi_{\mathcal{K},L}=B_{\mathcal{K}}\Biggl\{-w_{\mathcal{K},+;-}\Phi_++w_{\mathcal{K},+;+}\Phi_- 
+F_{\mathcal{K}}Z_{\mathcal{K}}\Psi_{-\mathcal{K},-}\\ -\dfrac{F_{\mathcal{K}}}{v}\sum\limits_{k'}\mathcal{P}\Biggl[\frac{\rho_2^*(\mathcal{K},-k')}{\mathcal{K}-k'}\Psi_{k',+} -\frac{\rho_1^*(\mathcal{K},-k')}{\mathcal{K}+k'}\Psi_{k',-}\Biggr] \Biggr\},
\label{LHESD}
\end{multline}
where 
\begin{align}
\rho_1(k,k')&=\left(w_{k,+;+}^*w_{k',+;+}+w_{k,+;-}^*w_{k',+;-}\right)/F_k,\\
\rho_2(k,k')&=\left(w_{k,+;-}^*w_{-k',+;+}^*-w_{k,+;+}^*w_{-k',+;-}^*\right)/F_k\,.
\label{rho}
\end{align}

Straightforward calculations show that the wave functions $\Psi_{\mathcal{K},R(L)}$ satisfy orthogonality relations
\begin{equation}
\langle \Psi_{\mathcal{K},R(L)}|\Psi_{\mathcal{K}',R(L)} \rangle=\delta_{\mathcal{K},\mathcal{K}'},\quad \langle \Psi_{\mathcal{K},R}|\Psi_{\mathcal{K}',L} \rangle=0,
\end{equation}
and the amplitude $B_{\mathcal{K}}$ is
\begin{equation}
B_{\mathcal{K}}=\frac{1}{F_{\mathcal{K}}\sqrt{Z_{\mathcal{K}}^2+(L/2v)^2}}
= \frac{1}{\sqrt{\left(E-\varepsilon_0-\Sigma_{\mathcal{K}}\right)^2+\gamma_{\mathcal{K}}^2}}\,,
\label{AmplitudeB}
\end{equation}
with $\gamma_{\mathcal{K}}=\dfrac{LF_{\mathcal{K}}}{2v}$. 

The functions $\rho_{1,2}$ have an important property:
\begin{equation}
\rho_1(k,k)=1,\quad \rho_2(k,-k)=0\,,
\label{rho12-0}
\end{equation}
which follows from the symmetry relations for the tunneling matrix~(\ref{w_symmetry}). Due to this property, the wave functions  $\Psi_{\mathcal{K},R(L)}$ have the following asymptotic behavior 
\begin{align}
\Psi_{\mathcal{K},R}\Big|_{x\to \pm \infty} \simeq e^{\mp i\phi_{\mathcal{K}}}\, \Psi_{\mathcal{K},+}\,,\\
\Psi_{\mathcal{K},L}\Big|_{x\to \pm \infty} \simeq e^{\pm i\phi_{\mathcal{K}}}\, \Psi_{\mathcal{K},-}\,,
\label{asymptotics}
\end{align}
where $\phi_{\mathcal{K}}$ is the phase that the wave function acquires when an electron passes the defect,
\begin{equation}
\tan\phi_{\mathcal{K}}= \frac{\gamma_{\mathcal{K}}}{E-\varepsilon_0 - \Sigma_{\mathcal{K}}}\,.
\end{equation}
Thus, the composite wave functions $\Psi_{\mathcal{K},R(L)}$ exactly correspond to the definition of the right- and left-moving states at infinity.

\begin{figure}
\includegraphics[width=0.9\linewidth]{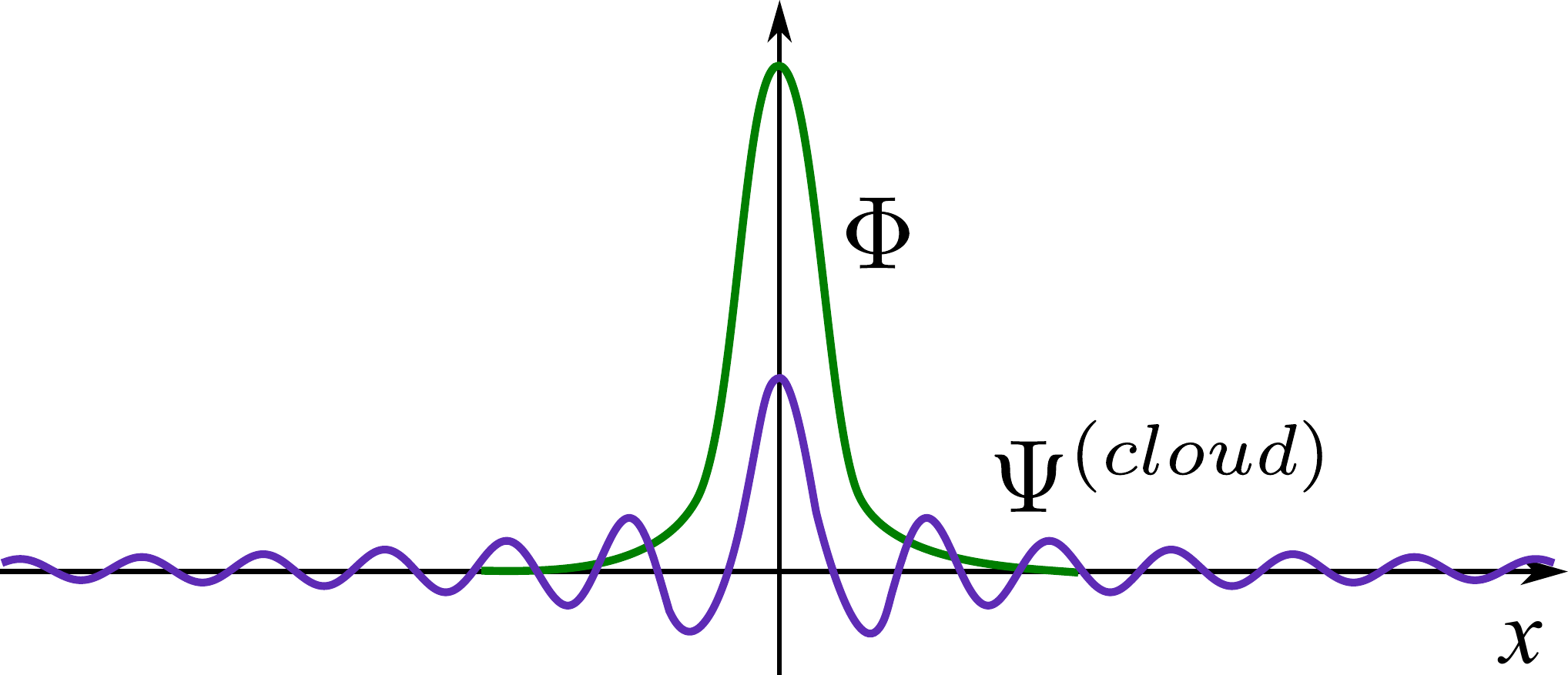}
 \caption{Sketch of the spatial dependence of the wave function components: the short-scale component that is the bound state $\Phi$, and the long-scale component that is the cloud formed by the left- and right-moving conventional HESs.} 
\label{fig_3}
\end{figure}

Now it is interesting to clarify how the composite wave functions are arranged at a finite distance from the defect. Equations~(\ref{RHESD}) and (\ref{LHESD}) show that $\Psi_{\mathcal{K},R(L)}$ contain three components: \\ 
1) a short-scale component localized at the defect, \\
2) a long-scale component extending far away from the defect and vanishing at infinity,\\
3) a propagating component defined by the asymptotics~(\ref{asymptotics}).
The spatial arrangement of the short-scale and long-scale components along the edge is schematically shown in Fig.~\ref{fig_3}.

The short-scale component is composed of bound states that are localized directly near the defect. For the right-moving states
\begin{equation}
\Psi_{\mathcal{K},R}^{(bound)}=B_{\mathcal{K}}\left(w_{\mathcal{K},+;+}^*\Phi_++w_{\mathcal{K},+;-}^*\Phi_-\right)\,,
\end{equation}
and similarly for the left-moving states.

The propagating component is formed by the third terms and the nonzero asymptotic part of the fourth term at $|x|\to \infty$ in Eqs.~(\ref{RHESD}) and~(\ref{LHESD}). For the right-moving states
\begin{equation}
\Psi_{\mathcal{K},R}^{(prop)}=B_{\mathcal{K}}F_{\mathcal{K}}\left[Z_{\mathcal{K}}\Psi_{\mathcal{K},+} +\dfrac{1}{v}\sum\limits_{k'}\mathcal{P}\frac{1}{\mathcal{K}-k'}\Psi_{k',+}\right],
\end{equation}
and similarly for the left-moving ones.

The long-scale component, which we call the cloud, is formed by the remaining part of the fourth terms in Eqs~(\ref{RHESD}) and (\ref{LHESD}). So, in the right-moving composite state, the cloud is described as follows:
\begin{align} 
\Psi_{\mathcal{K},R}^{(cloud)}=&\Psi_{+}^{(cloud)}+\Psi_{-}^{(cloud)}\nonumber\\
 & =\frac{B_{\mathcal{K}}F_{\mathcal{K}}}{v}\sum\limits_{k'}\mathcal{P}\frac{\rho_1(\mathcal{K},k')-1}{\mathcal{K}-k'}\Psi_{k',+}\nonumber \\
 &+\frac{B_{\mathcal{K}}F_{\mathcal{K}}}{v}\sum\limits_{k'}\mathcal{P}\frac{\rho_2(\mathcal{K},k')}{\mathcal{K}+k'}\Psi_{k',-}\,.
\label{cloud} 
\end{align}
The cloud is seen to consist of both Kramers partners of the conventional HESs.

The dependence of the cloud component of the wave function on the coordinate along the edge can be estimated using the asymptotics of the integrals in Eq.~(\ref{cloud}). This is easy to do, since matrix elements $w_{k,\sigma;\lambda}$, as functions of $k$, have no singularity and vanish at infinity. In addition, according to Eq.~(\ref{rho12-0}) the functions $\rho_1(\mathcal{K},k')-1$ and $\rho_2(\mathcal{K},k')$ are equal to zero in the points where the denominator is zero. Therefore, the integrands are regular functions and we have 
\begin{equation}
\Psi_{\pm}^{(cloud)}\Bigl|_{|x|\to \infty} \propto b_{\pm}(\mathcal{K},K_c) \frac{e^{iK_c x}}{x}-b_{\pm}(\mathcal{K},-K_c) \frac{e^{-iK_c x}}{x}\,,
\label{cloud_asymptotics}
\end{equation}
where $K_c$ is a cutoff momentum corresponding to the energy above which $w_{k,\sigma;\lambda}$ drops as shown in Fig.~\ref{fig_2}. The value of $K_c$ is determined by the band gap (more precisely, by a slightly higher energy) and equals approximately $K_c\approx |M|/v$. The function $b_{\pm}(\mathcal{K},K_c)$ is a four-rank spinor that depends on two arguments. The explicit expression for $b_{\pm}(\mathcal{K},q)$ is rather cumbersome, but it is important that $b_{\pm}$ is not zero, and its dependence on $\mathcal{K}$ is determined by the matrix elements $w_{k,\sigma;\lambda}$ and functions $\widetilde{\Psi}_{k,\sigma}$. More detail analysis shows that $\Psi_{\pm}^{(cloud)}$ can be roughly approximated as 
\begin{equation}
\Psi_{\pm}^{(cloud)}\Bigl|_{|x|\to \infty} \propto  \widetilde{b}_{\pm}(\mathcal{K}) \frac{\sin(K_cx)}{x}\,.
\label{cloud_asymptotics1}
\end{equation}

The energy dependence of the cloud amplitude is determined mainly by the factor $B_{\mathcal{K}}$ defined by Eq.~(\ref{AmplitudeB}). The amplitude has a resonance at the energy $E^{(res)}=\varepsilon_0+\Sigma_{\mathcal{K}}$. The resonance energy is shifted by the self-energy $\Sigma_{\mathcal{K}}$ from the bound state energy, as usually in the Fano-Anderson model~\cite{mahan2013many}. The width of the resonance $\gamma_{\mathcal{K}}$ is determined by all components of the tunneling matrix, therefore the participation of the spin-flip transitions increases its width. 

Of great interest is the fact that in the vicinity of the defect there is a fairly wide cloud of the conventional HESs with opposite Kramers indexes. In a sense, it could be said that due to the tunnel coupling of HESs and a defect, spin flipping or backscattering of conventional HESs occurs, but the ``backscattered'' component with the flipped spin disappears at infinity. 

The amplitude of the cloud with the flipped spin can be quite large. Asymptotically, it is estimated as
\begin{equation}
\Psi_{+\to -}^{(cloud)}\sim B_{\mathcal{K}}\frac{\sqrt{L}}{\pi v} \left(\overline{w}_{++} \overline{w'}_{+-}-\overline{w'}_{++}\overline{w}_{+-}\right)^*\overline{\widetilde{\Psi}}_{-}(y)\,\frac{\sin{K_c x}}{x}\,,
\end{equation}
where $\overline{w}_{+\pm}$ is an averaged value of $w_{k,+;\pm}$ over the interval $[-K_c, K_c]$,  $\overline{w'}_{+\pm}$ is an averaged $k$-derivative of $w_{k,+;\pm}$, and $\overline{\widetilde{\Psi}}_{-}$ is averaged $\widetilde{\Psi}_{k,-}$. 
In the resonance, the cloud amplitude is
\begin{equation}
\Psi_{+\to -}^{(cloud)}\Bigg|_{res}\! \sim \frac{\left(\overline{w}_{++} \overline{w'}_{+-}-\overline{w'}_{++}\overline{w}_{+-}\right)^*}{|w_{\mathcal{K},+;+}|^2+|w_{\mathcal{K},+;-}|^2}\, \overline{\widetilde{\Psi}}_{-}(y)\,\frac{\sin{K_c x}}{x}\,.
\end{equation}

It is clearly seen that the maximum amplitude of the cloud is determined by the factor, which depends not only on the magnitude of the tunneling matrix elements, but on their derivatives with respect to $k$. Particularly, the cloud disappears if the tunneling matrix does not depend on $k$. If we evaluate this factor using the data of Fig.~\ref{fig_2}, it turns out to be about 0.2. 

Thus, the cloud of the spin-flipped HESs is large enough to produce quite noticeable effects at a finite distance from the defect. In particular, nontrivial effects can arise in a system of many defects located near the edge. In the next section we show that defects can interact with each other through the edge states at a large distance exceeding very much the radius of the localization of the bound states.

\section{Coupling between defects through the edge states}\label{Sec3}

In a system of several defects located near the edge, a long-range perturbation of the edge states produced by each defect affects the bound states located at other defects. Thus, interaction between defects becomes possible, even if they are located at a large distance from one another, exceeding the characteristic length of their direct coupling, which is determined by the overlap of their wave functions. The idea of this effect, in a sense, stems from two well-known effects: the configuration interaction of a localized state and a continuum~\cite{PhysRev.124.1866}, and the RKKI indirect exchange interaction of magnetic moments~\cite{PhysRev.96.99}. In this section we study this mechanism of indirect coupling for two defects, which allows one to find out main effects of this interaction.

\begin{figure}
\includegraphics[width=0.95\linewidth]{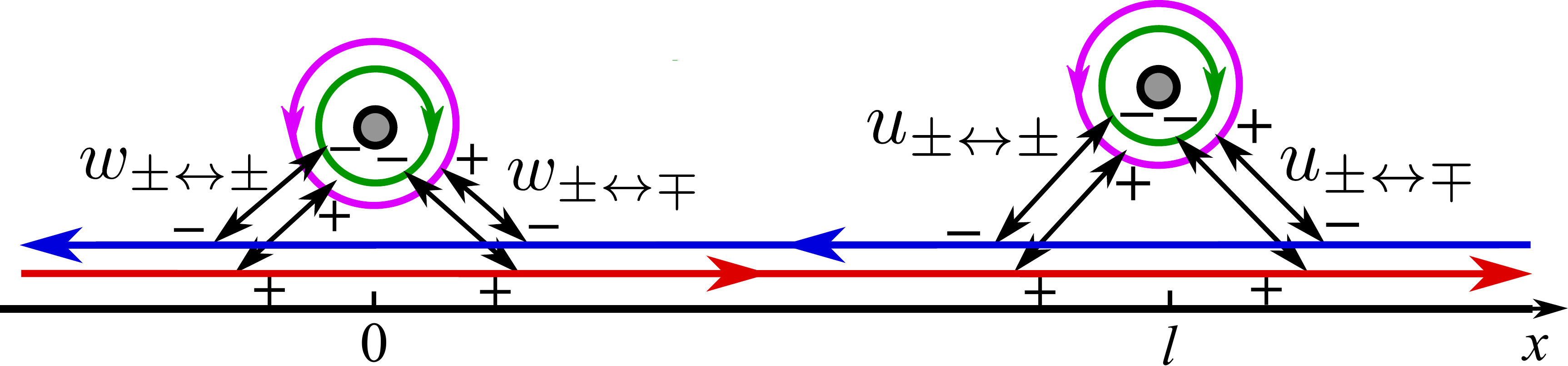}
 \caption{Indirect coupling between two defects through HESs.} 
\label{fig_4}
\end{figure}

Consider two, in the general case, different defects, located near the edge at a distance $l$ from one another along the edge, Fig.~\ref{fig_4}. For simplicity, we assume that the distance $l$ is sufficiently large, so that the overlap of the wave functions of states localized at different defects is negligible, and that each defect has only one energy level.

The Hamiltonian of the system is
\begin{multline}
H=\sum\limits_{k,\sigma}|k,\sigma\rangle \varepsilon_{k,\sigma}\langle k,\sigma| + \sum\limits_{\lambda}|\lambda\rangle \varepsilon_1\langle \lambda|+ \sum\limits_{\mu}|\mu\rangle \varepsilon_2\langle \mu| \\+ \sum\limits_{k,\sigma,\lambda}\!\left(|k,\sigma\rangle w_{k,\sigma;\lambda}\langle \lambda|+ h.c.\right)+ \sum\limits_{k,\sigma,\mu}\!\left(e^{-i k l}|k,\sigma\rangle u_{k,\sigma;\mu}\langle \mu|+ h.c.\right)\,,
\label{Two_defect_Hamiltonian}
\end{multline}
where $|\lambda\rangle$ and $|\mu\rangle$ are Kramers pairs of the bound states at the different defects with the energy levels $\varepsilon_1$ and $\varepsilon_2$, and $w_{k,\sigma;\lambda}$ and $u_{k,\sigma;\mu}$ are matrix elements describing the tunnel coupling of the corresponding defect and the HESs. The factor $e^{-i k l}$ takes into account the displacement of the defects relative to each other by a distance $l$ along the axis $x$. This form of the coupling Hamiltonian implies that the tunneling matrix for each defect are calculated in the coordinate system centered on this defect. In the Hamiltonian~(\ref{Two_defect_Hamiltonian}) we neglect the direct tunnel coupling between the bound states $|\lambda \rangle$ and $|\mu \rangle$, assuming that the distance $l$ far exceeds the characteristic length of the localization of their wave functions, which is of the order $\sqrt{|B/M|}$ in the BHZ model.

\subsection{Wave functions of combined HESs}\label{Sec_CHES2D}
Now we find the eigenfunctions of the Hamiltonian~(\ref{Two_defect_Hamiltonian}). This problem is solved in the same way as it was done for a single defect in Sec.~\ref{HESDs}, but the calculations are more complicated and cumbersome. The basic idea of the calculations with some details is given in Appendix~\ref{App2}.

Results of these calculations are as follows. There are two Kramers conjugate eigenfunctions describing right- and left-moving composite HESs. The wave function of a right-moving HES composed of two bound states and conventional HESs reads
\begin{multline}
\Psi_{\mathcal{K},R}=C_{\mathcal{K},R}\Biggl\{\sum_{\lambda}\beta_{\lambda}\Phi_{\lambda}+\sum_{\mu}\gamma_{\mu}X_{\mu}\\ 
+ Z_{\mathcal{K}} \mathcal{G}_+(\mathcal{K},\mathcal{K})\Psi_{\mathcal{K},+}+\sum_{\sigma}\sum_k \mathcal{P}\frac{\mathcal{G}_{\sigma}(\mathcal{K},k)}{E-\sigma v k}\Psi_{k,\sigma}\Biggr\},
\label{RHES2D}
\end{multline}
where $\Phi_{\lambda}$ and $X_{\mu}$ are the wave functions of the bound states located at different defects and $C_{\mathcal{K},R}$ is the normalization constant
\begin{equation}
C_{\mathcal{K},R}=\frac{1}{|\mathcal{G}_+(\mathcal{K},\mathcal{K})|\sqrt{Z_{\mathcal{K}}^2+\dfrac{L^2}{4v^2}}}\,.
\label{AmplitudeC}
\end{equation}

Though equations for the wave function and the normalization constant are similar in form to the corresponding equations in the case of single defect, there are the following important differences.

First, the function $Z_{\mathcal{K}}$, which largely determines the resonance energy, has now a more complicated form
\begin{equation}
Z_{\mathcal{K}}=\frac{\Delta_1 \Delta_2-|\Sigma_3|^2-|\Sigma_4|^2}{\Delta_1 F_2+\Delta_2 F_1+2\mathrm{Re}[\Sigma_3 F_3^*+\Sigma_4 F_4^*]}\,.
\label{Z_K_2}
\end{equation}
where 
\begin{equation}
\Delta_{1,2} =E-\varepsilon_{1,2}-\Sigma_{1,2}\,
\end{equation}
is the energy difference between the eigenenergy $E$ of the state and the resonant energy of the respective defect, if it is considered as isolated. The quantities $\Sigma_{1,2}$ are the corresponding self-energies of the isolated defects defined by Eqs.~(\ref{Sigma1}) and (\ref{Sigma2}), and $\Sigma_{3,4}$ are new characteristic energies that appear in the two-defect system. They are given by Eqs.~(\ref{Sigma3}) and~(\ref{Sigma4}). Four quantities $F_{1, 2, 3, 4}$ are given by Eqs.~(\ref{F1})-(\ref{F4}) in Appendix~\ref{App2}.
 
Second, the function $\mathcal{G}_{\sigma}(\mathcal{K},k)$ of two arguments appears instead of function $\rho_{1, 2}(\mathcal{K},k)$. It is defined as
\begin{equation}
\mathcal{G}_{\sigma}(\mathcal{K},k)=\sum_{\lambda}\beta_{\mathcal{K},\lambda}w_{k,\sigma;\lambda}+e^{-ikl}\sum_{\mu}\gamma_{\mathcal{K},\mu}u_{k,\sigma;\mu}\,,
\end{equation}
where the functions $\beta_{\mathcal{K},\sigma}$ and $\gamma_{\mathcal{K},\sigma}$ are given in Appendix~\ref{App2}.

The function $\mathcal{G}_{\sigma}(\mathcal{K},k)$ plays an important role, since it defines the asymptotics of the wave functions of the right- and left-moving composite states:
\begin{align}
\Psi_{\mathcal{K},R}\simeq & C_{\mathcal{K},R} \left[Z_{\mathcal{K}}-\frac{iL}{2v}\mathrm{sgn}(x)\right]\mathcal{G}_+(\mathcal{K},\mathcal{K})\Psi_{\mathcal{K},+}\,,\label{asymptotics_2dR}\\
\Psi_{\mathcal{K},L}\simeq & C_{\mathcal{K},L} \left[Z_{\mathcal{K}}+\frac{iL}{2v}\mathrm{sgn}(x)\right]\mathcal{G}_-(\mathcal{K},-\mathcal{K})\Psi_{\mathcal{K},+}\,.\label{asymptotics_2dL}
\end{align}

The wave function of the left-moving composite states differs from Eq.~(\ref{RHES2D}) by the obvious replacement of the signs of $\mathcal{K}$ and $\sigma$.

\subsection{Spectrum of the wave function amplitude}
Of greatest interest is the study of the spectral dependence of the wave function amplitude in the region where the defects are located, since it is this quantity that substantially depends on the interactions we are considering. First of all, it is important to study the situation when the defects are identical. In this case, in the absence of interaction between the defects, one can expect that the amplitude will have a resonance similar to the resonance of a single defect. In this section, we show that in fact the spectrum completely changes and in a very unusual way depends on the distance between defects.

\begin{figure}
\includegraphics[width=1.\linewidth]{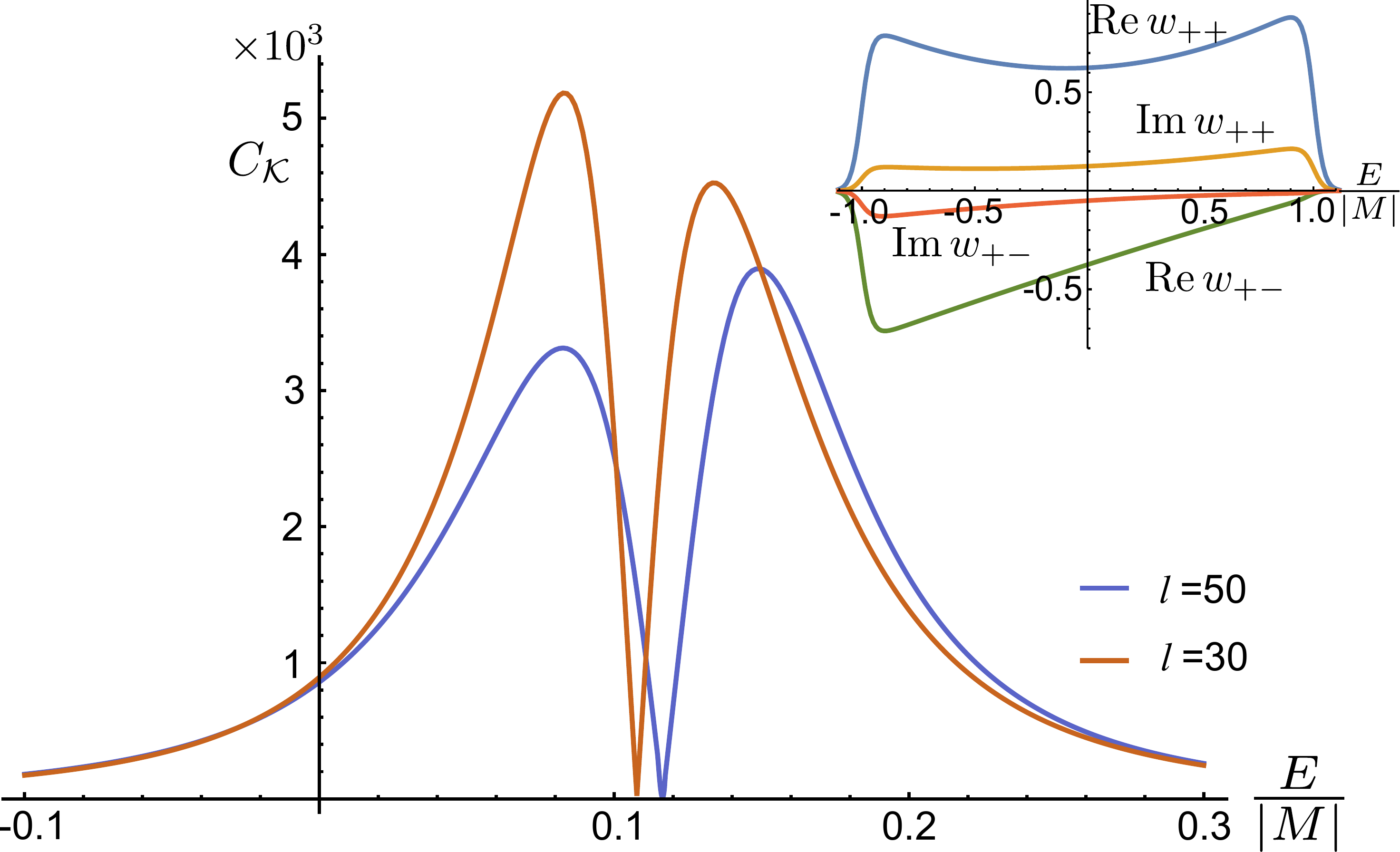}
 \caption{Amplitude $C_{\mathcal{K},R}$ of the helical edge state coupled to two defects as a function of the energy for two distances $l$ between the defects. The energy is normalized to $|M|$ and the distance is normalized to $\sqrt{B/M}$. Inset shows the tunneling matrix elements used in the calculation as a function of the energy.} 
\label{fig_5}
\end{figure}

If the defects are identical, the above equations are somewhat simplified due to the fact that $u_{k,\sigma;\mu} = w_{k,\sigma;\mu}$ and $\varepsilon_1=\varepsilon_2=\varepsilon_0$. Direct calculations of the amplitude $C_{\mathcal{K},R}$ as a function of the energy $E$ with using Eq.~(\ref{AmplitudeC}) lead to results shown in Fig.~\ref{fig_5} for the parameters used in the calculation of the tunneling matrix of Fig.~\ref{fig_2}. Similar results were obtained also for a wide range of the model parameters.

Analytical analysis shows that the general form of the spectrum of $C_{\mathcal{K},R}$ does not change substantially with varying the tunneling matrix. It is only important that $w_{k,\sigma;\lambda}$ has no singularities as a function of $k$ and vanishes outside the band gap. The main feature of the spectrum is that there are two peaks of resonances, which are generally asymmetric. Their position, form and asymmetry change with varying $w_{k,\sigma;\lambda}$ and distance $l$ between the defects. This conclusion is confirmed by numerical calculations for a wide range of parameters.

The origin of the main features of the amplitude spectrum can be understood from the analysis of the factor $[Z^2+(L/2v)^2]^{-1/2}$ in Eq.~(\ref{AmplitudeC}), which plays a key role. The amplitude is roughly approximated by the following expression
\begin{equation}
C_{\mathcal{K}}\propto \frac{1}{\sqrt{Z_{\mathcal{K}}^2F_{\mathcal{K}}^2+\gamma^2}}\,,
\label{AmplitudeCsimpl}
\end{equation}
which is quite similar to Eq.~(\ref{AmplitudeB}) for the case of a single defect. But now $Z_{\mathcal{K}}$ has the form
\begin{equation}
Z_{\mathcal{K}}=\frac{\Delta^2-\Sigma_{34}^2}{2F_{\mathcal{K}}(\Delta + W_{34})}\,,
\label{Z_K_simpl}
\end{equation}
where two important quantities are introduced, $\Sigma_{34}$ and $W_{34}$, which characterize the indirect coupling between defects. 

The quantity $\Sigma_{34}$ is defined as
\begin{equation}
\Sigma_{34}^2=|\Sigma_3|^2+|\Sigma_4|^2\,,
\end{equation}
which resembles in form a self-energy function, if we look at Eqs.~(\ref{Sigma3}) and (\ref{Sigma4}), but refers to two coupled defects, since it contains products of the matrix elements of both defects and the distance between them. Another characteristic energy is
\begin{equation}
W_{34}=\mathrm{Re}[\Sigma_3 F_3^*+\Sigma_4 F_4^*]\,,
\end{equation}
which also is determined by the products of the matrix elements $w_{k,\sigma;\lambda}$ and $u_{k,\sigma;\mu}$, and distance $l$.

Thus $C_{\mathcal{K}}$ reads
\begin{equation}
C_{\mathcal{K}}\propto \frac{|\Delta + W_{34}|}{\sqrt{(\Delta^2-\Sigma_{34}^2)^2+2\gamma^2(\Delta + W_{34})}}\,.
\label{AmplitudeCsimpl1}
\end{equation}
Comparison with the numerical calculation carried out using Eq.~(\ref{AmplitudeC}) shows that Eq.~(\ref{AmplitudeCsimpl1}) correctly describes the position of the resonances, and the factor $\mathcal{G}_+(\mathcal{K},\mathcal{K})$ affects the shape outside the peaks, in many cases significantly.

To better understand how the characteristic energies $\Sigma_{34}$ and $W_{34}$ affect the amplitude spectrum, we consider a simplified case when $\gamma \ll|\Sigma_{34}|$, which really takes place, as will be seen later. In this case, it is clear that the resonances arise when $\Delta_{1,2}^{(res)}\approx \pm|\Sigma_{34}|$, which corresponds to the energy $E_{1,2}^{(res)}=\varepsilon_0 + \Sigma_{\mathcal{K}} \pm |\Sigma_{34}|$, where $\Sigma_{\mathcal{K}}$ is defined by Eq.~(\ref{selfenergy0}). Thus, $\Sigma_{34}$ describes the shift of the resonances one relative to the other. The energy $ W_{34}$ makes the shape of the resonances asymmetric, in particular, asymmetrically changes their height and width.

Finally, we note that the fact that $C_{\mathcal{K}}$ vanishes at the energy $E=\varepsilon_0+\Sigma_{\mathcal{K}}-W_{34}$ does not mean that the wave function also vanishes. The matter is that the expression in braces in Eq.~(\ref{RHES2D}) has a singularity at this point, so that the wave function remains finite. This is most easily shown by the example of asymptotic behavior, Eqs.~(\ref{asymptotics_2dR}),~(\ref{asymptotics_2dL}). Using Eqs.~(\ref{AmplitudeC}) and (\ref{Z_K_simpl}) it is easy to see that at the point $\Delta+W_{34}=0$, $\Psi_{\mathcal{K},R}$ does not vanish.

\subsection{Discussion}
Since $\Sigma_{34}$ determines the energy splitting of the resonances of isolated defects, this quantity can be interpreted as a self-energy function of the indirect coupling between the defects, although this term may not be very precise. Therefore, it is interesting to find out how $\Sigma_{34}$ depends on the distance between the defects. 

\begin{figure}
\includegraphics[width=0.9\linewidth]{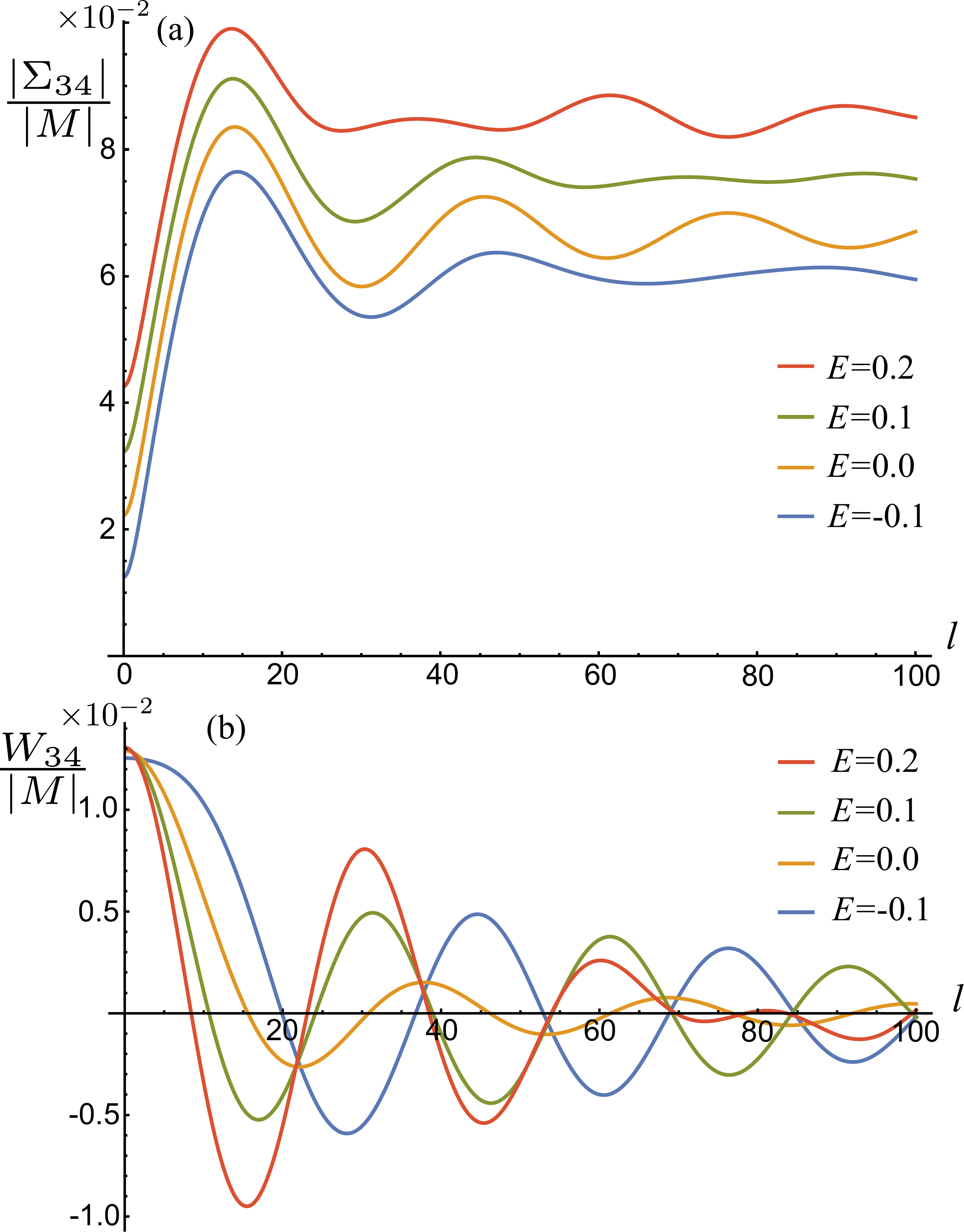}
 \caption{(a) The self-energy of the indirect configuration interaction $\Sigma_{34}$ and (b) the energy $W_{34}$, which determines the asymmetry of the resonances, as functions of the distance $l$ between the defects for different values of the energy $E$. For better viewing, the lines in the panel (a) are shifted upwards by 0.01 in series with increasing energy. The energy is normalized to $|M|$, the distance is normalized to $\sqrt{B/M}$.} 
 \label{fig_6}
\end{figure}

With this goal, we should turn to Eqs.~(\ref{Sigma3}) and (\ref{Sigma4}) that define $\Sigma_3$ and $\Sigma_4$. Since in both equations the integrands have a singularity at $vk'=E$ and contain the exponential factors $\exp(\pm i\mathcal{K}l)$, we can expect that $\Sigma_3$ and $\Sigma_4$, as functions of the energy, have two oscillating components, one of which oscillates with the wave vectors $\mathcal{K}$ and the other with $K_c$. The relative contribution of both components depends on the specific form of $w_{k,\sigma;\lambda}$ as functions of $k$. Numerical calculations carried out with using the tunneling matrix $w_{k,\sigma;\lambda}$ shown in the inset of Fig.~\ref{fig_5} lead to the results presented in Fig.~\ref{fig_6}(a).
Qualitatively similar results were obtained for other models of $w_{k,\sigma;\lambda}$ we considered.

It is interesting that $\Sigma_{34}$ varies with the distance $l$ quite differently than the cloud component of the wave function of an isolated defect. The energy $\Sigma_{34}$ oscillates with the distance, approaching a finite constant value in the limit of large $l$, while the amplitude of the cloud tends to zero as Eq.~(\ref{cloud_asymptotics}) shows. In addition, the form of the oscillations is also different. This is due to the fact that the cloud is strongly changed in the presence of two defects. In this case, the composition of the HESs forming the cloud changes radically since new types of transitions appear that also perturb the HESs. These are the transitions between the defects through the HESs. Therefore, the cloud in a two-defect system is not just a superposition of clouds of isolated defects. Our analysis shows that the limiting value of $\Sigma_{34}$ at $l \to \infty$ can be estimated as
\begin{equation}
 \Sigma_{34}\sim \pi^2 (|\overline{w}_{+,+}|^2+|\overline{w}_{+,-}|^2)(|\overline{u}_{+,+}|^2+|\overline{u}_{+,-}|^2)\,,
\end{equation} 
where the line over $w$ and $u$ means the averaging over $k$ in the band gap.

Of course, the distance up to which the indirect coupling acts is really limited by phase decoherence processes that were not taken into account. This is clear from the fact that the characteristic energies $\Sigma_{3, 4}$ substantially depend on the phase shift that the HESs acquire between the defects. This phase shift is described factors $\exp[\pm i \mathcal{K}l]$ in Eqs.~(\ref{Sigma3}) and~(\ref{Sigma4}). Decoherence processes add a random phase that destroys the long-range coupling of defects.

Another characteristic energy of the indirect coupling $W_{34}$, that determines the asymmetry of the resonances, also oscillates with the distance $l$ but tends to zero at infinity, Fig.~\ref{fig_6}b. Asymptotically, at $l\gg 1$, $W_{34}$ is approximated as
\begin{multline}
 W_{34}\sim -4\pi (|\overline{w}_{+,+}|^2+|\overline{w}_{+,-}|^2)(|\overline{u}_{+,+}|^2+|\overline{u}_{+,-}|^2)\\ \times \frac{K_c\sin K_cl \sin\mathcal{K}l+\mathcal{K}\cos K_cl \cos \mathcal{K}l}{(K_c^2-\mathcal{K}^2)l} \,.
\end{multline}

\begin{figure}
\includegraphics[width=0.9\linewidth]{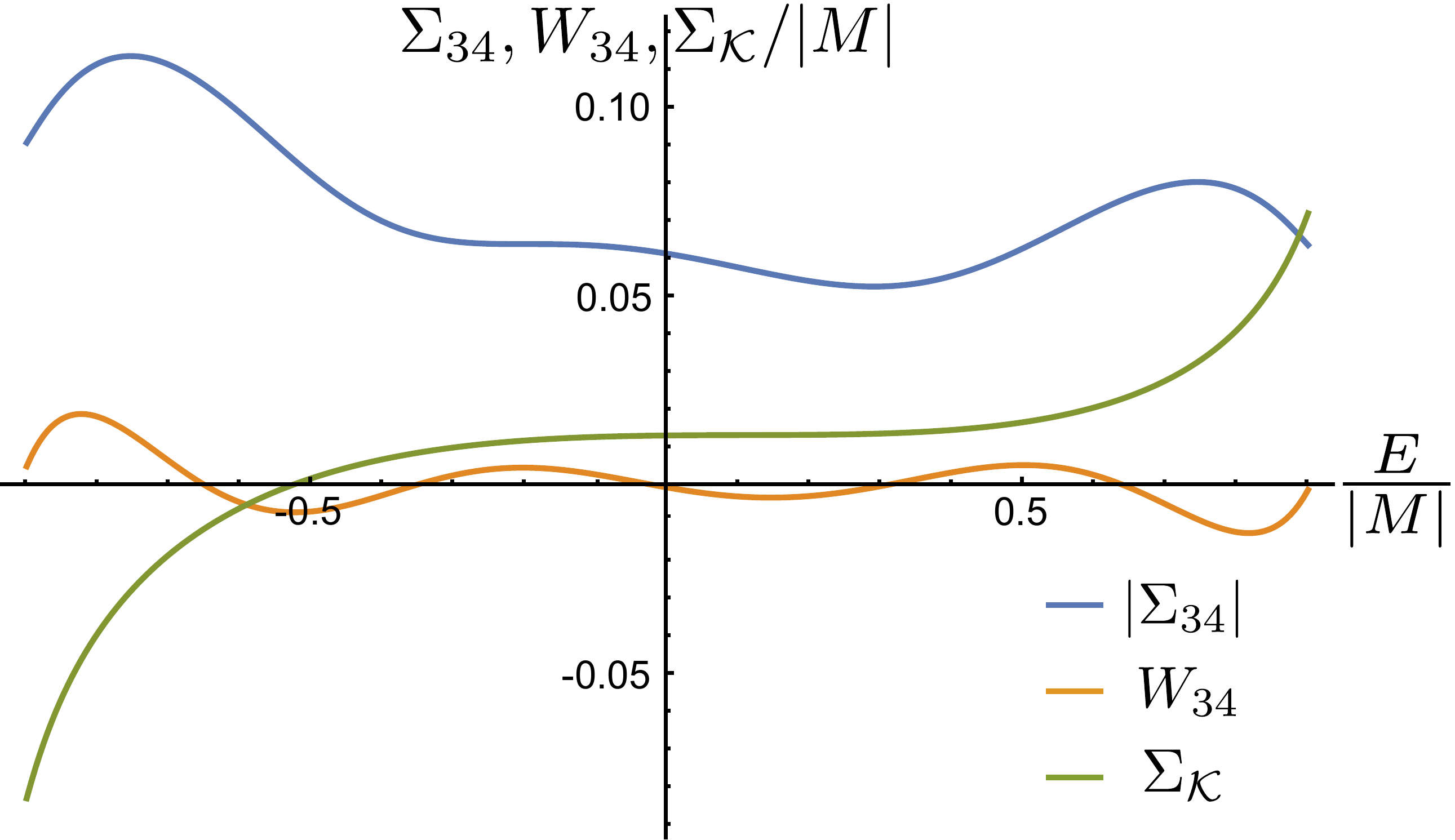}
\caption{(a) The indirect coupling energies $\Sigma_{34}$ and $W_{34}$ as functions of the energy $E$ of the state. The distance between defects is $l=50\sqrt{B/M}$, the energy is normalized to $|M|$. For comparison, the self-energy $\Sigma_{\mathcal{K}}$ of a single defect coupled to HESs is also shown.} 
\label{fig_7}
\end{figure}

The indirect coupling energies $\Sigma_{34}$ and $W_{34}$ depend also on the energy $E$ of the state. This dependence is shown in Fig.~\ref{fig_7} in the case of tunneling matrix of Fig.~\ref{fig_5}. It is seen that the energy dependence of $\Sigma_{34}$ and $W_{34}$ is smooth on the scale of $\gamma$, as we supposed.

In addition, Figs.~\ref{fig_6} and~\ref{fig_7} demonstrate that the self-energy component $\Sigma_{34}$ really far exceeds $W_{34}$. The significance of the indirect coupling energies can be estimated by comparing them with the self-energy $\Sigma_{\mathcal{K}}$ of a single defect coupled to the HESs, which is also shown in Fig.~\ref{fig_7}. As can be seen, $\Sigma_{34}$ exceeds $\Sigma_{\mathcal{K}}$ for reasonable values of the tunneling matrix elements and distance $l$.

\section{Conclusion}\label{Sec4}

We have studied the electronic states that are formed in a 2D TI as a result of the tunnel coupling of HESs and bound states localized at nonmagnetic point defects, in the general case when the axial spin symmetry is broken due to SOI. 

Like conventional HESs, the composite HESs are classified as moving to the right and left, but differ significantly in their electronic structure in an energy region near the resonances associated with the levels of the bound state. A composite HES with an energy $E$ is composed not only of the conventional HES with the wave number $E/v$, which falls on the defect from infinity and then goes to the opposite infinity, and the bound states on the defect. It contains also a set of the conventional HESs with wave vectors in a wide range of energy that form a cloud around the defect with a quite large amplitude near the resonance. The cloud extends over a considerable distance from the defect. Its amplitude decreases with distance asymptotically as $1/x$. 

Of great interest is the fact that in systems with broken spin symmetry, the cloud consists of both Kramers partners of the conventional HESs. This means that, for example, a right-moving composite HES contains a cloud of the conventional HESs with the Kramers index (in other words, spin structure) of the left-moving conventional HESs. Since at infinity, the spinor structure of composite and conventional right-moving HESs is the same, one can say that due to the tunnel coupling to the defect, a spin flip or backscattering occurs around the defect. The cloud of spin-flipped states exists only when the axial spin symmetry is broken. If there is no SOI, the cloud, of course, also exists, but it is composed only of the HESs with the same spin as the wave at infinity.

The presence of the clouds is interesting in the following aspects. First, the clouds can affect the electron-electron interaction and scattering of electrons with energy near the resonances. This problem requires a separate study. Secondly, although the cloud disappears at infinity, it can create nontrivial effects at a finite distance from the defect. One of these effects has been studied here. 

The effect arises in a system of many or several defects located near the edge. The defects can interact with each other through the edge states that they perturb. Since the perturbation of the HESs produced by each defect is extended over a large distance, an indirect coupling of the bound states at different defects occurs even if their wave functions do not overlap.

The indirect coupling has been studied for a system of two defects coupled through the HESs. The most striking effect occurs when the defects are identical. The indirect coupling leads to a splitting of the resonance of the isolated defects into two peaks, even if the distance between them significantly exceeds the localization length of the bound states. The magnitude of the splitting can be unexpectedly large and reach tenths of the band gap under realistic conditions. It is determined by the self-energy of the indirect coupling $\Sigma_{34}$, which depends in an unusual way on the distance between the defects. The splitting energy oscillates with increasing the distance and tends asymptotically to a constant value. The oscillations reflect an oscillating structure of the cloud component of the perturbed HESs in the two-defect system. 

Another effect of the indirect coupling is an asymmetry of the resonances, which vary in height and width. This feature is described by another characteristic energy $W_{34}$, which also oscillates with the distance, but tends to zero with removing the defects from one another. Of course, at extremely large distance the indirect coupling disappears because of phase decoherence processes not taken into account in this work.

The effects of the indirect coupling may turn out to be most interesting for topologically nontrivial materials in which scattering is suppressed.

\begin{acknowledgments}
The author thanks A. A. Sukhanov for numerical calculations of the tunnel matrix elements shown in Fig. 2. This work was supported by the Russian Science Foundation, project No.~16-12-10335.

\end{acknowledgments}

\appendix
\section{Tunneling Hamiltonian}\label{App1}

Here we propose a Hamiltonian describing the tunnel coupling between HESs and bound states.

Total Hamiltonian of a bounded 2D system containing a defect can be written in the form
\begin{equation}
H=H_{bulk}+U(y)+V(x,y-d)\,,
\label{Htot}
\end{equation}
where $H_{bulk}$ is the Hamiltonian of 2D TI, $U(y)$ is the Hamiltonian describing the presence of an edge at $y=0$, and $V(x,y-d)$ is the potential of a nonmagnetic defect located at $x=0, y=d$. To be specific we can consider $U(y)$ as an infinite wall described by a step function $U(y)=U\Theta(-y)$ with $U\to \infty$.

Following the Bardeen method~\cite{PhysRevLett.6.57}, we divide the system into two subsystems A and B coupled to each other: 
\begin{equation}
H = H_A + H_B + W,
\label{H_T}
\end{equation}
where $H_A=H_{bulk}+U(y)$ describes the bulk with the boundary and $H_B=H_{bulk}+V(r)$ describes the defect in the unbounded TI, with $r$ being the radial coordinate with respect to the defect. Eigenfunctions of $H_A$ are HESs $|k,\sigma\rangle$ and eigenfunctions of $H_B$ are the bound states $|n,\lambda\rangle$ at the defect. 

Equations~(\ref{Htot}) and (\ref{H_T}) clearly show that the tunneling Hamiltonian is $W=-H_{bulk}$. As $H_{bulk}$ we can take, for example, the BHZ Hamiltonian.

Thus, the tunneling matrix elements are estimated as
\begin{equation}
w_{k,\sigma;n,\lambda}=-\langle k,\sigma|H_{bulk}|n,\lambda\rangle = -\varepsilon_n\langle k,\sigma|n,\lambda\rangle + \langle k,\sigma|V|n,\lambda\rangle\,,
\label{tunnel_matrix}
\end{equation}
where $\varepsilon_n$ is the bound state energy with the quantum number $n$ added for generality. 

In Sec.~\ref{HESDs} we present the tunneling matrix elements calculated straightforwardly in the frame of the BHZ model with the SOI due to the bulk inversion asymmetry.

The BHZ model presents the electronic states in the basis $\left(|e\uparrow\rangle,|h\uparrow\rangle,|e\downarrow\rangle,|h\downarrow\rangle\right)^T$, where $|e\rangle$ and $|h\rangle$ denote the electron and hole band states with spin up and down. In materials with broken bulk inversion symmetry, the Hamiltonian reads~\cite{Bernevig1757,doi:10.1143/JPSJ.77.031007} 
\begin{equation}
H=\begin{pmatrix}
 M\!-\!B k^2 & Ak_+ & 0 & -\Delta\\
 Ak_- & -M\!+\!B k^2 & \Delta & 0\\
 0 & \Delta & M\!-\!B k^2 & -Ak_-\\
 -\Delta & 0 & -Ak_+ & -M\!+\!B k^2 
\end{pmatrix}\,,
\label{HamiltonianBHZ}
\end{equation}
where $M$, $A$, $B$ are well-known parameters of the model, $\Delta$ is the SOI parameter, $\mathbf{k}$ is the momentum, $k_{\pm}=k_x\pm ik_y$. 

The wave functions of the HESs are calculated by the method described in Sec.~\ref{HESs}, and the bound-state wave functions are calculated in the case of the short-range potential $V(\mathbf{r})$ using the method developed previously~\cite{doi:10.1002/pssr.201409284,SABLIKOV20161}. 

\section{Helical edge states coupled to two defects}\label{App2}

This section provides details of the calculation of the wave functions for the system studied in Sec.~\ref{HESDs}, where we study a system of two defects coupled to HESs. The system is described by the Hamiltonian~(\ref{Two_defect_Hamiltonian}). 

The eigenfunctions of the Hamiltonian~(\ref{Two_defect_Hamiltonian}) are constructed in the form
\begin{equation}
 \Psi=\sum\limits_{k',\sigma'}A_{k',\sigma'}|k',\sigma'\rangle + \sum\limits_{\lambda'}B_{\lambda'}|\lambda'\rangle + \sum\limits_{\mu'}C_{\mu'}|\mu'\rangle\,.
\end{equation}

Coefficients $A_{k,\sigma}$, $B_{\lambda}$ and $C_{\mu}$ are determined by the stationary Schr\"odinger equation, from which it follows that $A_{k,\sigma}$ is related to $B_{\lambda}$ and $C_{\mu}$ by the equation
\begin{equation}
A_{k,\sigma}=\frac{1}{E-\varepsilon_{k,\sigma}}
\left[\sum_{\lambda}B_{\lambda}w_{k,\sigma;\lambda}+e^{-ikl}\sum_{\mu}C_{\mu}u_{k,\sigma;\mu}\right]\,,
\end{equation}
with
\begin{equation}
\frac{1}{E-\varepsilon_{k,\sigma}}=\mathcal{P}\frac{1}{E-\varepsilon_{k,\sigma}}+Z_{\mathcal{K}}\delta_{k,\sigma\mathcal{K}}\,.
\end{equation}
Coefficients $B_{\lambda}$ and $C_{\mu}$ are determined by a homogeneous system of four linear equations with matrix
\begin{equation}
\mathfrak{M}=
\begin{pmatrix}
 M_1 & 0 & M_3 & M_4\\
 0 & M_1 & -M_4^* & M_3^*\\
 M_3^* & -M_4 & M_2 & 0\\
 M_4^* & M_3 & 0 & M_2 
\end{pmatrix}\,,
\end{equation}
where
\begin{align}
M_{1,2} & =-\Delta_{1,2}+Z_{\mathcal{K}}F_{1,2}\,,\\
M_{3,4} & = \Sigma_{3,4}+Z_{\mathcal{K}}F_{3,4}\,,
\end{align}
and
\begin{equation}
\Delta_{1,2} =E-\varepsilon_{1,2}-\Sigma_{1,2}\,.
\end{equation}
Here
\begin{align}
\Sigma_1 &\!=\!\sum\limits_{k'}\mathcal{P}\frac{|w_{k',+;+}|^2+|w_{k',+;-}|^2}{E-vk'},\label{Sigma1} \\
\Sigma_2 &\!=\!\sum\limits_{k'}\mathcal{P}\frac{|u_{k',+;+}|^2+|u_{k',+;-}|^2}{E-vk'},\label{Sigma2}\\
\Sigma_3 &\!=\!\sum\limits_{k'}\mathcal{P}\frac{u_{k',+;+}w_{k',+;+}^*e^{-ik'l}+u_{k',+;-}^*w_{k',+;-}e^{ik'l}}{E-vk'},\label{Sigma3}\\
\Sigma_4 &\!=\!\sum\limits_{k'}\mathcal{P}\frac{u_{k',+;-}w_{k',+;+}^*e^{-ik'l}-u_{k',+;+}^*w_{k',+;-}e^{ik'l}}{E-vk'},\label{Sigma4}
\end{align}
\begin{align}
F_1 &=|w_{\mathcal{K},+;+}|^2+|w_{\mathcal{K},+;-}|^2,\label{F1}\\
F_2 &=|u_{\mathcal{K},+;+}|^2+|u_{\mathcal{K},+;-}|^2,\label{F2}\\
F_3 & = u_{\mathcal{K},+;+}w_{\mathcal{K},+;+}^*e^{-i\mathcal{K}l}+u_{\mathcal{K},+;-}^*w_{\mathcal{K},+;-}e^{i\mathcal{K}l},\label{F3}\\
F_4 & = u_{\mathcal{K},+;-}w_{\mathcal{K},+;+}^*e^{-i\mathcal{K}l}-u_{\mathcal{K},+;+}^*w_{\mathcal{K},+;-}e^{i\mathcal{K}l}.\label{F4}
\end{align}

The requirement that the determinant of the matrix $\mathfrak{M}$ be equal to zero gives the equation
\begin{equation}
M_1 M_2 - |M_3|^2-|M_4|^2=0\,
\end{equation}
that allows one to determine $Z_{\mathcal{K}}$. The analysis of this equation shows that there is a single root which is presented by Eq.~(\ref{Z_K_2}) in Sec.~\ref{Sec_CHES2D}.

As a result of fairly simple but cumbersome calculations of the coefficients $A_{k,\sigma}$, $B_{\lambda}$, and $C_{\mu}$, we arrive at Eq.~(\ref{RHES2D}) for the wave function which is presented in Sec.~\ref{Sec_CHES2D}. In this equation the wave function is expressed in terms of auxiliary functions $\beta_{\mathcal{K},\sigma}$ and $\gamma_{\mathcal{K},\sigma}$: 
\begin{align}
\beta_{\mathcal{K},+}\!=&\left(|M_3|^2+|M_4|^2\right)w_{\mathcal{K},+;+}^*\!\!-\!M_1\left(M_3^*u_{\mathcal{K},+;+}+M_4u_{\mathcal{K},+;-}^*\right)e^{i\mathcal{K}l},\\
\beta_{\mathcal{K},-}\!=& \left(|M_3|^2+|M_4|^2\right)w_{\mathcal{K},+;-}^*\!\!+\!M_1\left(M_4^*u_{\mathcal{K},+;+}-M_3^*u_{\mathcal{K},+;-}^*\right)e^{i\mathcal{K}l},\\
\gamma_{\mathcal{K},+}\!=& M_1\left(M_1u_{\mathcal{K},+;+}^*e^{i\mathcal{K}l}+M_4w_{\mathcal{K},+;-}^*-M_3^*w_{\mathcal{K},+;+}^*\right)\,,\\
\gamma_{\mathcal{K},-}\!=& M_1\left(M_1u_{\mathcal{K},+;-}^*e^{i\mathcal{K}l}-M_3w_{\mathcal{K},+;-}^*-M_4^*w_{\mathcal{K},+;+}^*\right)\,.  
\label{beta_gamma}
\end{align}

\vspace{1cm}
~
\bibliography{edge_def}

\end{document}